\documentclass[preprint,showpacs,preprintnumbers,amsmath,amssymb]{revtex4}
\usepackage{graphicx,epsfig,subfigure,latexsym,makeidx,amsmath}
\usepackage{color}
\usepackage{float}
\usepackage[colorlinks,linkcolor=blue,anchorcolor=blue,citecolor=green,urlcolor=blue]{hyperref}
\usepackage[english]{babel}
\usepackage{amsmath,amssymb,amsthm}
\usepackage{bm}
\usepackage[mathscr]{eucal}

\begin{document}
	\newcommand \nn{\nonumber}
	\newcommand \fc{\frac}
	\newcommand \lt{\left}
	\newcommand \rt{\right}
	\newcommand \pd{\partial}
	\newcommand \e{\text{e}}
	\newcommand \hmn{h_{\mu\nu}}
	\newcommand{\PR}[1]{\ensuremath{\left[#1\right]}} 
	\newcommand{\PC}[1]{\ensuremath{\left(#1\right)}} 
	\newcommand{\PX}[1]{\ensuremath{\left\lbrace#1\right\rbrace}} 
	\newcommand{\BR}[1]{\ensuremath{\left\langle#1\right\vert}} 
	\newcommand{\KT}[1]{\ensuremath{\left\vert#1\right\rangle}} 
	\newcommand{\MD}[1]{\ensuremath{\left\vert#1\right\vert}} 

	
	\title{ Exact black hole solutions in gravity with a background Kalb-Ramond field }
	
	\preprint{}

	\author{Jia-Zhou Liu, Shan-Ping Wu, Shao-Wen Wei, Yu-Xiao Liu \footnote{Corresponding author. E-mail: liuyx@lzu.edu.cn}}

	\affiliation{$^{1}$Lanzhou Center for Theoretical Physics, Key Laboratory of Theoretical Physics of Gansu Province, and Key Laboratory of Quantum Theory and Applications of MoE, Lanzhou University, Lanzhou, Gansu 730000, China,\\
		$^{2}$Institute of Theoretical Physics $\&$ Research Center of Gravitation, Lanzhou University, Lanzhou 730000, People's Republic of China}

	\begin{abstract}

In this work, we derive exact solutions for  four-dimensional static spherically symmetric black holes and three-dimensional rotating black holes within a Lorentz-violating gravity theory. In this framework, Lorentz symmetry is spontaneously broken when a nonminimally coupled Kalb-Ramond tensor field acquires a nonzero vacuum expectation value. Building upon these solutions, we investigate the thermodynamic properties of the black holes using the Iyer-Wald  formalism. Our findings reveal that the standard first law of thermodynamics and the Smarr relation remain valid for black holes in the presence of the Kalb-Ramond field.
	
\end{abstract}	

\maketitle
\newpage
\tableofcontents
\newpage

\section{INTRODUCTION }
Several theories of quantum gravity have been proposed~\cite{Birrell:1982ix,Maldacena:1997re,Aharony:1999ti,Gubser:1998bc,Alfaro:1999wd,Alfaro:2001rb,Rovelli:1989za}, but lots of quantum gravity effects can be observed only at the Planck scale ($\sim 10^{19}$ GeV), which is beyond the reach of current experimental capabilities. Some of these quantum gravity theories assume that Lorentz symmetry might be broken in the gravitational IR regime, which corresponds to our presently accessible low-energy scales.

Stemming from the concept of spontaneous Lorentz symmetry breaking  in string theory~\cite{Kostelecky:1988zi}, the Standard Model extension was introduced as a suitable framework for violating Lorentz symmetry and  was widely studied~\cite{Colladay:2001wk,Kostelecky:2000mm,Kostelecky:2001mb,Colladay:2009rb,Berger:2015yha,Carroll:1989vb,Andrianov:1994qv,Andrianov:1998wj,Lehnert:2004hq,BaetaScarpelli:2012kt,Brito:2013npa}. In this context, the bumblebee model serves as a straightforward toy model~\cite{Kostelecky:1989jw,Kostelecky:1988zi,Kostelecky:2003fs}, with the simplest self-interacting vector field $B_\mu$ called bumblebee vector which couples with gravity nonminimally. The bumblebee vector field $B_{\mu}$ obtains a nonzero vacuum expectation value (VEV) through a specified potential, resulting in the spontaneous breaking of Lorentz symmetry in the gravitational sector.

In the context of Bumblebee gravity, Casana et al. obtained an exact solution for a static spherically symmetric spacetime~\cite{Casana:2017jkc}. Subsequently, Maluf et al. derived an (A)dS-Schwarzschild-like solution by relaxing the vacuum conditions~\cite{Maluf:2020kgf}, while Xu et al. identified new classes of static spherically symmetric Bumblebee black holes with a nonzero temporal component of the 
bumblebee vector field and investigated their thermodynamic properties as well as observational implications~\cite{Xu:2022frb}. Ding et al. obtained a slowly rotating Bumblebee black hole solution and a BTZ-like black hole solution~\cite{Ding:2019mal,Ding:2023niy}. Moreover, several new black hole solutions were recently  proposed within the framework of bumblebee gravity~\cite{Liu:2024axg,Ding:2024qrf}. Different properties 
of these   black holes were extensively explored in various works~\cite{Filho:2022yrk,Kuang:2022xjp,Oliveira:2018oha,Liu:2022dcn,Gomes:2018oyd,Kanzi:2019gtu,Gullu:2020qzu,Oliveira:2021abg,Kanzi:2021cbg,Ovgun:2018ran,Sakalli:2023pgn,Mangut:2023oxa,AraujoFilho:2024ykw}.

In another Lorentz violation model, the Kalb-Ramond field $B_{\mu\nu}$, which arises from
the bosonic spectrum of string theory~\cite{Kalb:1974yc}, is a self-interacting second-rank antisymmetric tensor field. When the tensor field’s nonzero VEV is nonminimally coupled with the gravity sector, it leads to spontaneous Lorentz symmetry violation. This coupling results in the ground state of a physical quantum system being characterized by nontrivial VEVs.
Similarly, several black hole solutions have also been proposed and studied in the context of the Kalb-Ramond  field~\cite{Do:2020ojg,Liu:2024oas,Duan:2023gng,Yang:2023wtu,Lessa:2019bgi,Kumar:2020hgm,Liu:2024lve}. The corresponding properties of the black holes were also studied~\cite{Du:2024uhd,Guo:2023nkd,Junior:2024ety,Jumaniyozov:2024eah,Ma:2024ets,al-Badawi:2024pdx, Zahid:2024hyy, Wei:2014dka, Zahid:2024ohn,Atamurotov:2022wsr,Lessa:2020imi,Maluf:2021ywn,Liu:2024wpa,Jha:2024xtr,Fathi:2025byw,Ghosh:2023xes,Filho:2023ycx,AraujoFilho:2024ctw,Nair:2024xdb}.

It is worth noting that some previous studies have treated the variation of the nonminimal coupling term between the Kalb–Ramond field and the scalar curvature, $\xi_1 B^{\mu\nu}B_{\mu\nu}R$, as effectively transforming into $\mp \xi_1 b^2 R$ in the vacuum. Here, $b_{\mu\nu}$ denotes the vacuum expectation value of $B_{\mu\nu}$, and $b^{\mu\nu}b_{\mu\nu} =\mp b^{2} $. As a result, they argued that this term could be absorbed into the action via a redefinition of the gravitational coupling constant, making it appear as if the term vanishes from the action. However, this interpretation is misleading. In fact, the variation of $\xi_1 B^{\mu\nu}B_{\mu\nu}R$ with respect to the metric is not equivalent to the variation of $\mp \xi_1 b^2 R$. Therefore, the contribution of this term to the gravitational field equations has been overlooked in such treatments. As will be discussed in more detail around Eq.~\eqref{BBR} –\eqref{bbR}, this nonminimal coupling plays a crucial role in our analysis.

 In this paper, we will investigate the effects of correctly incorporating this term on black hole solutions. Our goal is to obtain black hole solutions under more general model, where the Kalb-Ramond field is nonminimally coupled to gravity. Furthermore, we analyze the corresponding black hole thermodynamics using the Iyer-Wald covariant phase space formalism.

This paper is organized as follows. In Sec.~II, we provide a review of the Kalb-Ramond model in the context of spontaneous Lorentz symmetry breaking. Sec.~III presents the static spherically symmetric black hole solution with a background Kalb-Ramond field, both with and without a cosmological constant. In Sec.~IV, we extend this solution to the case of a three-dimensional rotating black hole. The thermodynamic properties of these black holes are analyzed in Sec.~V using the Iyer-Wald covariant phase space formalism. Finally, Sec.~VI presents a summary and discussion of the work.
\section{THE KALB-RAMOND MODEL FOR SPONTANEOUS LORENTZ SYMMETRY BREAKING  }
Let us consider the Einstein-Hilbert action nonminimally coupled to a self-interacting Kalb–Ramond field in the form~\cite{Altschul:2009ae}
\begin{eqnarray}
	S&=&\int d^{D}x\sqrt{-g}\bigg[\frac{1}{2\kappa}(R-2\Lambda)-\frac{1}{12}H^{\mu\nu\rho}H_{\mu\nu\rho}-V(B^{\mu\nu}B_{\mu\nu})+\frac{1}{2\kappa}(\xi_{1}B^{\mu\nu}B_{\mu\nu}R \nonumber \\
	&&+\xi_{2}B^{\rho\mu}B^{\nu}{}_{\mu}R_{\rho\nu})\bigg],
	\label{S1}
\end{eqnarray}
where $\kappa=\frac{8\pi G}{c^4}$ is the gravitational coupling constant, $\Lambda$ represents the cosmological constant, and $\xi_{1}$ and $\xi_{2}$ are coupling constants governing the interaction between gravity and the Kalb-Ramond field.

The Kalb-Ramond field is a second-rank antisymmetric tensor field, denoted as $B_{\mu\nu}$, whose field strength is
\begin{equation}H_{\mu\nu\rho}\equiv\partial_{[\mu}B_{\nu\rho]}.\end{equation}
 The field strength $H_{\mu\nu\rho}$  
 remains invariant under a gauge
 transformation of $B_{\nu\rho}$
\begin{equation}B_{\nu\rho}\to B_{\nu\rho}+\partial_{\nu}\Lambda_{\rho}-\partial_{\rho}\Lambda_{\nu},\end{equation}
where $\Lambda_{\rho}$ is an arbitrary 1-form~\cite{Altschul:2009ae}.

Similar to the gravitational sector of the Standard Model extension~\cite{Colladay:2001wk,Kostelecky:2000mm,Kostelecky:2001mb,Colladay:2009rb,Berger:2015yha,Carroll:1989vb,Andrianov:1994qv,Andrianov:1998wj,Lehnert:2004hq,BaetaScarpelli:2012kt,Brito:2013npa}, we consider a self-interacting potential for the Kalb-Ramond field. We assume a potential of the general form  
\begin{equation}  
	V = V\left(B^{\mu\nu} B_{\mu\nu} \pm b^{2} \right)
\end{equation}  
with a nonzero VEV  $\langle B_{\mu\nu} \rangle = b_{\mu\nu}$. It follows that the VEV of the Kalb-Ramond field is determined by the condition $V(B^{\mu\nu} B_{\mu\nu} \pm b^{2})=0$, which implies the constraint 
\begin{equation}  
	b^{\mu\nu} b_{\mu\nu} = \mp b^{2}.
\end{equation}
 For convenience, we define
$X=B^{\mu\nu}B_{\mu\nu}\pm b^{2}$ and
$V'=\frac{\partial V}{\partial X}$, which will be used in the subsequent discussions.
As demonstrated in Ref.~\cite{Altschul:2009ae}, it is convenient to decompose the antisymmetric tensor as  
\begin{equation}  
	B_{\mu\nu} = {\tilde{E}}_{[\mu} v_{\nu]} + \epsilon_{\mu\nu\alpha\beta} v^{\alpha} \tilde{B}^{\beta},  
\end{equation}  
where $v^\mu$ is a timelike 4-vector. The background vectors $\tilde{E}$ and $\tilde{B}$ can be interpreted as a pseudo-electric field and a pseudo-magnetic field, respectively. These fields are spacelike, satisfying the orthogonality conditions  
\begin{equation}  
	\tilde{E}_{\mu} v^{\mu} = 0, \quad \tilde{B}_{\mu} v^{\mu} = 0.  
\end{equation}  
Consequently, the VEV of the Kalb-Ramond field leads to two independent background vectors, in contrast to the single vector generated by  the VEV of the 
bumblebee field.  In spherical coordinates $(t,r,
\theta,\phi)$, we assume that the vacuum configuration of the Kalb–Ramond field has only the following nonvanishing components:
\begin{equation}  
	b_{rt} = -b_{tr} = \tilde{E}(r).  
	\label{ansatz}  
\end{equation} 

We note that some previous works considered the term $ \xi_1 B^{\mu\nu} B_{\mu\nu} R $ in the action \eqref{S1} to transform into $ \mp \xi_1 b^2 R $ in vacuum, and thought that this term can be absorbed 
into the Einstein-Hilbert term through a redefinition of variables~\cite{Yang:2023wtu,Lessa:2019bgi,Duan:2023gng}. However, by explicitly varying the term $ B^{\mu\nu} B_{\mu\nu} R \sqrt{-g}$ with respect to $g^{\mu\nu}$, we obtain
\begin{eqnarray}
	\frac{\delta( B^{\alpha\beta}B_{\alpha\beta}R\sqrt{-g})}{\delta g^{\mu\nu}}&=&\big[g_{\mu\nu}\nabla^{2}(B^{\alpha\beta}B_{\alpha\beta})
	+B^{\alpha\beta}B_{\alpha\beta}G_{\mu\nu} \nonumber \\
	&-&2B^{\alpha}_{~\mu}B_{\nu}{}_{\alpha}R-\nabla_{\mu}\nabla_{\nu}(B^{\alpha\beta}B_{\alpha\beta})\big]\sqrt{-g}.
	\label{BBR}
\end{eqnarray}
While by varying the term $b^2R\sqrt{-g} $ with respect to $ g^{\mu\nu} $, we obtain
\begin{eqnarray}
	\frac{\delta( b^2R\sqrt{-g})}{\delta g^{\mu\nu}}= b^2G_{\mu\nu}\sqrt{-g}.
	\label{bbR}
\end{eqnarray}
When $B^{\mu\nu} B_{\mu\nu} = \mp b^2$, it is evident that Eq.~\eqref{BBR} is not equivalent to Eq.~\eqref{bbR}. This discrepancy underscores a critical issue: some previous studies have incorrectly argued that the term $\xi_{1}B^{\mu\nu}B_{\mu\nu}R$ can be absorbed into the Einstein–Hilbert action through a redefinition of variables. However, such a treatment fails to account for the fact that the variation of this term yields a nontrivial contribution to the gravitational field equations. As a result, the dynamical influence of this nonminimal coupling has been overlooked in those analyses.

 Varying the action with respect to $g^{\mu\nu}$ yields the gravitational field equations
\begin{equation}
	G_{\mu\nu}+\Lambda g_{\mu\nu}= \kappa T^{KR}_{\mu\nu} ,
	\label{modified}
\end{equation}
where 
\begin{equation}\begin{aligned}
		T_{\mu\nu}^{\mathrm{KR}}& =\frac{1}{2}H_{\mu\alpha\beta}H_{\nu}^{\alpha\beta}-\frac{1}{12}g_{\mu\nu}H^{\alpha\beta\rho}H_{\alpha\beta\rho}+\xi_{1}\big[\nabla_{\mu}\nabla_{\nu}(B^{\alpha\beta}B_{\alpha\beta})-g_{\mu\nu}\nabla^{2}(B^{\alpha\beta}B_{\alpha\beta})-B^{\alpha\beta}B_{\alpha\beta}G_{\mu\nu}  \\
		&+2B^{\alpha}_{\mu}B_{\nu}{}_{\alpha}R\big]+\xi_{2}\biggl[\frac{1}{2}g_{\mu\nu}B^{\alpha\gamma}B^{\beta}{}_{\gamma}R_{\alpha\beta}-B^{\alpha}{}_{\mu}B^{\beta}{}_{\nu}R_{\alpha\beta}-B^{\alpha\beta}B_{\nu\beta}R_{\mu\alpha}-B^{\alpha\beta}B_{\mu\beta}R_{\nu\alpha} \\
		&+\frac{1}{2}\nabla_{\alpha}\nabla_{\mu}\left(B^{\alpha\beta}B_{\nu\beta}\right)+\frac{1}{2}\nabla_{\alpha}\nabla_{\nu}\left(B^{\alpha\beta}B_{\mu\beta}\right)-\frac{1}{2}\nabla^{\alpha}\nabla_{\alpha}\left(B_{\mu}^{\gamma}B_{\nu\gamma}\right)-\frac{1}{2}g_{\mu\nu}\nabla_{\alpha}\nabla_{\beta}\left(B^{\alpha\gamma}B^{\beta}{}_{\gamma}\right)\bigg] \\
		&+4V^{\prime}(X)B_{\alpha\mu}B^{\alpha}{}_{\nu}-g_{\mu\nu}V(X).
\end{aligned}\end{equation}

Next, we will solve the above 
equations and give  static spherically symmetric black hole solutions under different scenarios.

\section{STATIC SPHERICALLY SYMMETRIC  BLACK HOLE SOLUTIONS}
We consider the following metric ansatz for a four-dimensional  static spherically symmetric spacetime:
\begin{equation}
	{d}{s}^{2}=-A(r) {dt}^{2}+S(r
	){dr}^{2}+r^{2} {~d}\Omega ^{2},
\end{equation}
where $d \Omega ^{2}={~d} \theta^{2}+ \sin ^{2} \theta {d} \varphi^{2}$.

The   ansatz of the VEV takes the form \eqref{ansatz} . Since $g^{\mu\alpha}g^{\nu\beta}b_{\mu\nu}b_{\alpha\beta} =-b^{2}, 
 $ combined with the metric ansatz, the pseudo-electric field $\tilde{E}$  can be written as
\begin{equation}\tilde{E}(r)=|b|\sqrt{\frac{A(r)S(r)}{2}}.\end{equation}
For convenience, we set $\ell_1=b^2\xi_1$ and $\ell_2=b^2\xi_2$ as  Lorentz-violating
parameters. By combining the spherically symmetric metric, we can obtain the specific field equations:
\begin{eqnarray}
	\frac{\left(2 \ell_1+\ell_2\right) A'(r)^2}{4 A(r) S(r)}-\frac{\left(2 \ell_1+\ell_2\right) A''(r)}{2 S(r)}+\frac{\left(2 \ell_1+\ell_2\right) A'(r) S'(r)}{4 S(r)^2}+\frac{\left(2 \ell_1+\ell_2\right) A'(r)^2}{4 A(r) S(r)}&& \nonumber \\
	-\frac{\left(4 \ell_1+\ell_2\right) A'(r)}{2 r S(r)}-\frac{\left(2 \ell_1-\ell_2+2\right) A(r)}{2 r^2 S(r)}+\frac{\left(\ell_1+1\right) A(r)}{r^2}&& \nonumber \\ 
	+\frac{\left(\ell_1+1\right) A(r) S'(r)}{r S(r)^2}-\Lambda  A(r)-V(X)A(r)-4b^2V^{\prime}(X)A(r)&=&0, \label{G1}\\
	\frac{\left(2 \ell_1+\ell_2\right) A''(r)}{2 A(r)}-\frac{\left(2 \ell_1+\ell_2\right) A'(r) S'(r)}{4 A(r) S(r)}-\frac{\left(2 \ell_1+\ell_2\right) A'(r)^2}{4 A(r)^2}&&\nonumber \\
	+\frac{\left(\ell_1+1\right) A'(r)}{r A(r)}-\frac{\left(\ell_1+1\right) S(r)}{r^2}+\frac{2 \ell_1-\ell_2+2}{2 r^2}&&\nonumber \\
	-\frac{\left(4 \ell_1+\ell_2\right) S'(r)}{2 r S(r)}+\Lambda  S(r)+V(X)S(r)+4b^2V^{\prime}(X)S(r)&=&0,\label{G2} \\
	\frac{\left(2 \ell_1+\ell_2-2\right) r^2 A'(r) S'(r)}{8 A(r) S(r)^2}+\frac{\left(2 \ell_1+\ell_2-2\right) r^2 A'(r)^2}{8 A(r)^2 S(r)} &&\nonumber \\
	-\frac{\left(2 \ell_1+\ell_2-2\right) r \left(r A''(r)+A'(r)\right)}{4 A(r) S(r)}+\frac{\left(2 \ell_1+\ell_2-2\right) r S'(r)}{4 S(r)^2}&&\nonumber \\
	+\Lambda  r^2+V(X)r^2+4b^2V^{\prime}(X)r^2&=&0.
	\label{G3}
\end{eqnarray}
\
Next, we will solve the above equations and give spherically symmetric black hole solutions under different scenarios.
\subsection{Case A: $V(X)=\frac{\lambda}{2}X^2$ and $ \Lambda=0$}
In the absence of the cosmological constant,  we impose the vacuum conditions  $V=0$ and $V'=0$~\cite{Casana:2017jkc,Majumdar:1999jd}.
An illustrative example of the potential satisfying these conditions is readily presented by a smooth quadratic form:
\begin{equation}
	V(X)=\frac{\lambda}{2}X^2,
	\label{VV}
\end{equation}
where $\lambda$ is a constant. This form is same as the  potential form of the Higgs field, and is also related to the mass structure of the theory~\cite{Kostelecky:1989jw}. In this case, the potential $V$ has no contribution to the field equations. Other choices of the potential, such as $V(X) \equiv \frac{\lambda}{2} X^n$ ($n\geq3$), also have no contribution to the field equations. Therefore, the solutions for these choices are consistent with the potential $ V(X) = \frac{\lambda}{2} X^2 $.
Substituting the expressions of   
$V$ in Eq.~\eqref{VV} the above 
condition into the field  Eqs.~\eqref{G1} - \eqref{G3}, the solution is given by:
\begin{eqnarray}
	A(r)&=&\frac{(1+\ell _1)}{1+\ell _1-\frac{\ell _2}{2}} -\frac{2 M}{r}, \nonumber \\
	S(r)&=&\frac{1}{A(r)}, \nonumber \\
	\tilde{E}(r)&=&{\frac{\sqrt{2}}{2}}|b|.
	\label{AS1}
\end{eqnarray}
It is worth noting that, since $b$ is a constant, the pseudo-electric field $\tilde{E}(r)$ also remains constant. In principle, Lorentz-violating parameter can take negative values, but it is constrained to be a small quantity~\cite{Yang:2023wtu}. Current bounds are
 \begin{eqnarray}
	-1.1\times 10^{-10} \leq \ell \leq 5.4\times 10^{-10}  .
\end{eqnarray}  By defining the parameter $\gamma = \frac{\ell_2}{2 + 2\ell_1}$, the metric function simplifies to 
 \begin{eqnarray}
A(r) = \frac{1}{1 - \gamma} -\frac{2 M}{r}.
 \end{eqnarray}
 It is straightforward to observe that this solution resembles the Kalb-Ramond black hole solution obtained by Yang et al.~\cite{Yang:2023wtu}. Furthermore, by performing a suitable rescaling of the coordinates, one can recover a black hole solution similar to the static, neutral, spherically symmetric solution proposed by Liu et al.~\cite{Liu:2024oas} within the framework of Kalb–Ramond gravity:
 \begin{eqnarray}
 	A(r)&=&1 -\frac{2(1-\gamma) M}{r}, \nonumber \\
 	S(r)&=&\frac{1-\gamma}{A(r)}.
 \end{eqnarray}

For the black hole solution corresponding to the metric in Eq.~\eqref{AS1}, the Kretschmann scalar takes the form
\begin{eqnarray}
	K&=&R_{\alpha\beta\delta\gamma}R^{\alpha\beta\delta\gamma} \nonumber \\ 
	&=&\frac{48 M^2}{r^6}-\frac{8\ell_2 M }{(1+\ell_1 - \frac{\ell_2}{2})r^5}+\frac{{\ell_2}^2  }{(1+\ell_1 - \frac{\ell_2}{2})^2 r^4}.
\end{eqnarray} 
This indicates that Lorentz-violating effects are intrinsic to the theory and cannot be removed by any coordinate transformation. The explicit breaking of Lorentz symmetry sets this framework apart from many other modified gravity theories. It is  evident that the metric functions approach to $A(r)\rightarrow\frac{1}{1 - \gamma}$ and $S(r)\rightarrow1 - \gamma$ at infinity. By performing the coordinate transformations $dt= \sqrt{1 - \gamma}\, d\hat{t}$ and $dr = \sqrt{1/(1 - \gamma)}\, d\hat{r}$, the asymptotic metric becomes
	\begin{eqnarray}
		ds^{2} = -d\hat{t}^{2} + d\hat{r}^{2} + \frac{1}{1- \gamma}\,\hat{r}^{2} d\Omega^{2}.
			\label{tr1}
	\end{eqnarray}
	This shows that the temporal and radial sectors coincide with those of Minkowski spacetime in spherical coordinates, while the angular part acquires a constant factor $1/(1-\gamma)$. And the Ricci scalar evaluates to
	\begin{eqnarray}
		R = -\frac{2\gamma}{(1 - \gamma)r^{2}}.
	\end{eqnarray}
	Hence, the spacetime is not asymptotically Minkowski. From another perspective, for asymptotically flat spacetimes one usually writes $r=\sqrt{x^{2}+y^{2}+z^{2}}$. As $r\to\infty$ along either timelike or lightlike directions, the metric takes the asymptotic form  
	\begin{equation}
		g_{\mu\nu} = \eta_{\mu\nu} + \mathcal{O}(r^{-1}),
	\end{equation}
	where $\eta_{\mu\nu}$ denotes the Minkowski metric.
	However, the asymptotic form of our metric can be written as
	\begin{eqnarray}
		ds^{2} = - d\hat t^{2} + (1 - \gamma)\,dr^{2} + r^{2} d\Omega^{2},
	\end{eqnarray}
	so that the deviation from Minkowski spacetime  in the radial component is exactly $\gamma$, i.e., a constant, rather than decaying as $\mathcal{O}(r^{-1})$. This shows that the spacetime is not asymptotically flat.

\subsection{Case B: $V(X)=\frac{\lambda}{2}X$ and $ \Lambda\neq0$}
Next, we extend our analysis to the case where the cosmological constant is nonzero. Our objective is to obtain an exact analytical black hole solution within this framework. The straightforward option for the potential can be a linear function~\cite{Duan:2023gng}:
\begin{equation}
	V(\lambda,X)=\frac{\lambda}{2}X,
	\label{L2}
\end{equation}
where $\lambda$ is a Lagrange multiplier field~\cite{Bluhm:2007bd}. The equation of motion obtained by varying with respect to $\lambda$ enforces the vacuum condition $X=0$, which implies $V=0$ for any field $\lambda$ on shell. In this way,  the construction effectively freezes motion about the potential minimum and hence permits an efficient extraction of the essential physics~\cite{Kostelecky:1989jw}. Unlike the vacuum condition in Case A, $V^{\prime}\neq0$ is valid when the field $\lambda$ is nonzero. Thus, the potential $V$ contributes to the field equations.
Combining the spherically symmetric metric and the form of the potential $V$, we can derive the specific field equations. Substituting the conditions given in Eqs.~\eqref{L2} and \eqref{la} back into the field equations~\eqref{G1}--\eqref{G3}, an exact analytical solution can be obtained, provided the coupling constants satisfy  
	\begin{equation}
		\lambda = \frac{(4\xi_{1}+\xi_{2})\Lambda}{\kappa \left(1-\ell_{1}-\tfrac{\ell_{2}}{2}\right)} ,
		\label{la}
	\end{equation}
	which follows directly from the field equations and therefore must hold whenever the latter are satisfied. 
	The exact analytical solution is give
\begin{eqnarray}
	A(r)&=&\frac{(1+\ell _1)}{1+\ell_1-\frac{\ell_2}{2}} -\frac{2 M}{r}-\frac{\Lambda }{3(1-\ell_1-\frac{\ell _2}{2})}r^2, \nonumber \\
	S(r)&=&\frac{1}{A(r)}.
\end{eqnarray}
It is easy to see that this solution is similar to the Schwarzschild-(A)dS-like black solution obtained by Yang et al.~\cite{Yang:2023wtu}. From this, we can see that when considering the nonminimal coupling of the Kalb-Ramond field with both the Ricci tensor and the scalar curvature, both contributions  affect the resulting black hole solutions.
We analyze the asymptotic behavior of the metric, taking into account both signs of the effective cosmological constant. After performing a coordinate transformation analogous to Eq.~\eqref{tr1}, the metric can be written in the asymptotic form:
\begin{equation}
	ds^{2} = -(1-\Lambda_e \hat{r}^2)\, d\hat t^{2} + \frac{1}{1-\Lambda_e \hat{r}^2}\, d\hat{r}^{2} + \frac{1}{1-\gamma}\, \hat{r}^{2} d\Omega^{2},
\end{equation}
where
\begin{equation}
	\Lambda_e = \frac{\Lambda}{(1-\gamma)(1-\ell_1 - \tfrac{\ell_2}{2})}.
\end{equation}
For $\Lambda_e>0$, the spacetime is asymptotically dS, with a cosmological horizon at $\hat{r} \sim 1/\sqrt{\Lambda_e}$. For $\Lambda_e<0$, the spacetime is asymptotically AdS.
The constant factor $1/(1-\gamma)$ in the angular sector corresponds to a uniform rescaling of the boundary sphere. Accordingly, the boundary conformal metric reads
\begin{equation}
	ds_{\rm bdry}^{2} = -d\hat\tau^{2} + \frac{1}{1-\gamma}\, d\Omega^{2}.
\end{equation}
where $\hat\tau$ denotes the time coordinate induced on the conformal boundary. Thus, up to this constant rescaling of the angular coordinates, the spacetime retains its asymptotic (A)dS character.

\subsection{Case C: $V(X)=\frac{\lambda}{2}X^2$ and $ \Lambda\neq0$}
In previous studies on gravity with a background Kalb-Ramond field and Bumblebee gravity, no analytical black hole solutions with a cosmological constant have been obtained for the potential $V(X) = \frac{\lambda}{2}X^2$. However, by considering the nonminimal coupling of the Kalb-Ramond field with the scalar curvature, we have obtained analytical solutions for a nonzero cosmological constant in this potential.

When $\ell_2 = -4\ell_1$, equivalently $\xi_2 = -4\xi_1$, and the potential is taken to be $V(X) = \frac{\lambda}{2} X^2$, the field equations \eqref{G1}–\eqref{G3} admit an exact analytical solution, given by:
\begin{eqnarray}
	A(r)&=&\frac{(1+\ell _1)}{1+3\ell _1} -\frac{2 M}{r}-\frac{\Lambda }{3(1+\ell_1)}r^2, \nonumber \\
	S(r)&=&\frac{1}{A(r)}.
\end{eqnarray}
Since we simultaneously consider the nonminimal couplings of the Kalb–Ramond field to both the Ricci tensor and the scalar curvature, the potential choice $V(X) = \frac{\lambda}{2}X^2$ admits a unified treatment of both cases with vanishing and nonvanishing cosmological constant. Moreover, in this case, the Lagrange multiplier field $\lambda$ no longer needs to account for the dependence on the cosmological constant.

\section{ROTATING BTZ-LIKE BLACK HOLE SOLUTION}
In the following,  we investigate stationary axial symmetric black hole solution in the three dimensional spacetime, within the framework of the Einstein–Hilbert action nonminimally coupled to a self-interacting Kalb–Ramond field. The stationary axial symmetric black hole metric in the three dimensional spacetime has the form
\begin{equation}
	{d}{s}^{2}=-A(r) {dt}^{2}+S(r
	){dr}^{2}+r^{2}[K(r) dt+{d}\varphi]^{2}.
	\label{g3}
\end{equation}
We consider the ansatz that the VEV of the Kalb–Ramond field takes the form \eqref{ansatz} and satisfies the condition $g^{\mu\alpha}g^{\nu\beta}b_{\mu\nu}b_{\alpha\beta} =-b^{2}
$. The Kalb–Ramond field is supposed as 

\begin{equation}
	b_{\mu \nu }=\left(
	\begin{array}{ccc}
		0 & -b \sqrt{\frac{A(r) S(r) }{2}} & 0 \\
		b \sqrt{\frac{A(r) S(r) }{2 }} & 0 & 0 \\
		0 & 0 & 0
		\label{b3}
	\end{array}
	\right).
\end{equation}
We consider the potential $V$ in Eq.~\eqref{VV} in Sec.~III Case B and the expression of $\lambda$ reads as
\begin{equation}
	\lambda=2\frac{(3\xi_1+\xi_2)\Lambda}{\kappa(1-\ell_1-\frac{\ell_2}{2})}.
	\label{la3}
\end{equation}
  Meanwhile, in order to get an analytic solution with rotation, we further consider that the two coupling constants have the following relationship $\xi_1=-\xi_2$. At this stage, we set $\ell=\frac {\xi_2 b^2}{2} $, and by substituting Eqs.~\eqref{g3}-\eqref{la3} into Eq.~\eqref{modified}, we finally obtain the corresponding black hole solution
\begin{eqnarray}
	A(r)&=&-M-\frac{\Lambda }{(1+{\ell})}r^2+\frac{j^2}{4(1+\ell)r^2}, \nonumber \\
	S(r)&=&\frac{1}{A(r)}, \nonumber \\
	K(r)&=&-\frac{j}{2r^2}.
\end{eqnarray} 
The $ g_{tt} $ component of the metric is given by
\begin{equation}
 g_{tt}=-A(r)+r^2 K(r)^2=M+\frac{\Lambda }{(1+{\ell})}r^2+\frac{\ell j^2}{4(1+\ell)r^2}.
\end{equation}
It is straightforward to observe that when $ \ell = 0 $, the solution is consistent with the rotating BTZ black hole solution~\cite{Banados:1992wn}. When $ j = 0 $, the solution reduces to a static BTZ-like black hole. The curvature of the static BTZ-like black hole is 
\begin{eqnarray}
R=\frac{6 \Lambda }{1+{\ell} }.
\end{eqnarray}
We can define the effective cosmological constant as $\Lambda_e = \frac{\Lambda}{1 + \ell}$, which satisfies the relation $R_{\mu\nu} = \Lambda_e g_{\mu\nu}$ at the boundary of the spacetime. This solution has two horizons: the Cauchy horizon at $r_-$ and the event horizon at $r_+$, both depending on the Lorentz-violating parameter $\ell$, which can be determined from the condition $A(r_{\pm}) = 0$:
\begin{equation}
		r_{\pm}=\sqrt{\frac{1}{-2\Lambda}}\left(\sqrt{(1+\ell) M \pm \sqrt{(1+\ell)^2 M^2 + \Lambda j^2}}\right).
\end{equation}
In general, an ergoregion appears when $g_{tt}>0$, which requires  
\[
(1+\ell)^{2}M^{2}-\Lambda \ell j^{2}\ge0 .
\]  
Solving $g_{tt}=0$, the ergosurface radii are obtained as  
\begin{equation}
	r_{\rm ergo}^{\pm}=\sqrt{\frac{1}{-2\Lambda}}
	\left(
	\sqrt{(1+\ell)M \pm \sqrt{(1+\ell)^{2}M^{2}-\Lambda \ell j^{2}}}
	\right).
\end{equation}
In the region $r_{\rm ergo}^{-} < r < r_{\rm ergo}^{+}$ we have $g_{tt}>0$. Since $\ell$ is a small parameter ($|\ell|\ll 1$) and $\Lambda<0$, it follows that $r_{\rm ergo}^{+} > r_{+}$ and $r_{\rm ergo}^{-} < r_{+}$, so the ergoregion extends over $r_{+} < r < r_{\rm ergo}^{+}$. Furthermore, the condition $(1+\ell)^{2} M^{2} - \Lambda \ell j^{2} \ge 0$ is automatically satisfied whenever the horizon exists, i.e., when $(1+\ell)^{2} M^{2} + \Lambda j^{2} \ge 0$.

Therefore, the Lorentz-violating parameter $\ell$ does not remove the ergoregion but modifies its size relative to the standard BTZ case. An ergoregion outside the event horizon $r_{+}$ always persists.

The Kretschmann scalar for this spacetime is given by
\begin{eqnarray}
	K=R_{\alpha\beta\delta\gamma}R^{\alpha\beta\delta\gamma}=\frac{12 \Lambda ^2}{(1+{\ell} )^2}=12{\Lambda_e} ^2,
\end{eqnarray}
which is a finite constant in the whole spacetime as like as the original BTZ black hole.
\section{THERMODYNAMICS}

Black hole thermodynamics is a key area in black hole studies, capturing semi-classical quantum gravitational effects in spacetime~\cite{Hawking:1975vcx, Gibbons:1976ue,Bardeen:1973gs,Hawking:1976de}. In our work, the Kalb–Ramond field is nonminimally coupled to gravity, leading to corrections that extend beyond those predicted by general relativity. This necessitates the construction of a modified thermodynamic framework to accurately describe the black hole solutions.
 To address these issues, we employ the Iyer-Wald formalism~\cite{Wald:1993nt,Iyer:1994ys,Iyer:1995kg}.

Consider the $D-$dimensional spacetime Lagrangian density for gravity and the Kalb-Ramond field, 
\begin{equation}
	\mathbf{L} = L \bm{\epsilon},
\end{equation}
where $\bm{\epsilon}$ denotes the spacetime volume form and $L$ corresponds to the Lagrangian in action \eqref{S1}. Varying the fields $\Phi \equiv \{g_{ab}, B_{ab}\}$ gives
\begin{equation}
	\delta\mathbf{L} = \mathbf{E}[\Phi]\delta\Phi + \mathrm{d}\mathbf{\Theta}[\Phi,\delta\Phi],
	\label{eq:LagrangianVariation}
\end{equation}
where $\mathbf{E}[\Phi]$ represents the bulk contribution from the variation (with  $\mathbf{E}[\Phi] =0$ corresponding to the field equations), and $\mathbf{\Theta}[\Phi,\delta\Phi]$ is the presymplectic potential.

It is clear that the action we consider is diffeomorphism invariant; under a field variation $\delta_\xi \Phi = \mathcal{L}_\xi \Phi$, the action remains unchanged. In other words, the Lagrangian density is a $D$-form, and its variation under $\delta_\xi \Phi = \mathcal{L}_\xi \Phi$ is given by
\begin{equation}
	\delta _{\xi}\mathbf{L}=\mathcal{L} _{\xi}\mathbf{L}=\xi \cdot \mathrm{d}\mathbf{L}+\mathrm{d(}\xi \cdot \mathbf{L})=\mathrm{d(}\xi \cdot \mathbf{L}).
\end{equation}
By applying the specific variation given in Eq.~\eqref{eq:LagrangianVariation}, we replace the variation of an arbitrary field with $\delta_\xi \Phi = \mathcal{L}_\xi \Phi$. This substitution yields
\begin{equation}
	\delta_\xi\mathbf{L} = \mathbf{E}[\Phi]\mathcal{L}_\xi\Phi + \mathrm{d}\mathbf{\Theta}[\Phi,\mathcal{L}_\xi\Phi],
\end{equation}
which, in turn, implies the identity
\begin{equation}
	\mathrm{d}(\xi \cdot \mathbf{L}) = \mathbf{E}[\Phi]\mathcal{L}_\xi\Phi + \mathrm{d}\mathbf{\Theta}[\Phi,\mathcal{L}_\xi\Phi].
\end{equation}
By invoking Noether’s second theorem, we conclude that the first term on the right-hand side is an exact form and vanishes on-shell. Consequently, the Noether current $ \textbf{J}_\xi $, which is conserved on-shell (i.e., $\mathrm{d}\mathbf{J}_\xi = 0$), is given by
\begin{equation}
	\mathbf{J}_\xi = \mathbf{\Theta}[\Phi, \mathcal{L}_\xi\Phi] - \xi \cdot \mathbf{L}.
	\label{eq:NoetherCurrent}
\end{equation}
 Furthermore, by the Poincaré lemma, this implies that at least locally there exists a $(D-2)$-form, called the Noether charge $\mathbf{Q}_\xi$, such that
\begin{equation}
	\mathbf{J}_\xi = \mathrm{d}\mathbf{Q}_\xi.
	\label{eq_JdQ}
\end{equation}
Building on this expression, taking the variation of the fields in Eq.~\eqref{eq:NoetherCurrent} yields
\begin{align}
	\delta \mathrm{d}\mathbf{Q}_{\xi}&=\delta [\mathbf{\Theta }(\Phi ,\mathcal{L} _{\xi}\Phi )]-\xi \cdot \delta \mathbf{L} \nonumber
	\\
	&=\delta [\mathbf{\Theta }(\Phi ,\mathcal{L} _{\xi}\Phi )]-\mathcal{L} _{\xi}[\mathbf{\Theta }[\Phi ,\delta \Phi ]]+d(\xi \cdot \mathbf{\Theta }[\Phi ,\delta \Phi ]).
\end{align}
This implies that
\begin{equation}
	\mathrm{d}\left( \delta \mathbf{Q}_{\xi}-\xi \cdot \mathbf{\Theta }[\Phi ,\delta \Phi ] \right) =\delta [\mathbf{\Theta }(\Phi ,\mathcal{L} _{\xi}\Phi )]-\mathcal{L} _{\xi}[\mathbf{\Theta }[\Phi ,\delta \Phi ]].
	\label{eq_dkxi}
\end{equation}
The right-hand side of this equality is precisely the symplectic current $\bm{\omega}[\delta\Phi,\mathcal{L}_\xi\Phi]$, which corresponds to the surface charge $H_{\xi}$ associated with the field variation $\delta_\xi \Phi = \mathcal{L}_\xi \Phi$. Specifically, it can be expressed as
\begin{equation}
	\delta H_{\xi} =\int_{S_{\infty}}{\delta \mathbf{Q}_{\xi}-\xi \cdot \mathbf{\Theta }[\Phi ,\delta \Phi ]}
	\label{eq_deltaH}.
\end{equation}
If $ \xi $ is a Killing vector field and $ \mathcal{L}_{\xi} \Phi = 0 $, then the symplectic current $\bm{\omega}( \delta \Phi, L_\xi \Phi)$ vanishes, leading to the simplification of Eq.~\eqref{eq_dkxi} to
\begin{equation}
	\mathrm{d}(\delta \textbf{Q}_\xi - \xi \cdot \mathbf{\Theta}[\Phi, \delta \Phi]) = 0.
	\label{Q2}
\end{equation}
For a stationary black hole with a bifurcate Killing horizon $S_h$ generated by $\xi_H$, we integrate this equation over a hypersurface $V_r $ that extends from the $S_h$ to another codimension-2 surface $ S_r $. Applying Stokes' theorem, the volume integral (for hypersurface) reduces to a surface integral, yielding
\begin{equation}
	\int_{S_r} (\delta\textbf{Q}_{\xi_H} - \xi_H \cdot \mathbf{\Theta}[\Phi, \delta \Phi]) - \int_{S_h} \delta \textbf{Q}_{\xi_H} = 0.
	\label{eq_FirstLawGen}
\end{equation}
Since the Killing vector $\xi_H$ vanishes on bifurcate surface $S_h$, its contribution to the second term simplifies. Notably, the properties of $\xi_H$ and $S_h$ allow us to express
\begin{equation}
	\int_{S_h}\delta\mathbf{Q}_{\xi_H} = T\delta S_W, 
\end{equation}
where $S_W$ is the Wald entropy. On the other hand, the integral over $S_r$ generally yields variations in physical quantities such as $\delta M$ and $\delta J$. Thus, Eq.~\eqref{eq_FirstLawGen} connects the entropy variation to changes in other thermodynamic quantities, corresponding to the first law of black hole thermodynamics.

In the above discussion , the conventional variation $\delta$ acts only on the fields while keeping the coupling parameters fixed. However, in many cases, the cosmological constant is treated as a variable pressure in the thermodynamics of AdS black holes, leading to an extended formulation of black hole thermodynamics~\cite{Kubiznak:2012wp,Dolan:2010ha,Dolan:2011xt,Kubiznak:2016qmn,Wei:2015iwa,Cai:2013qga,Kastor:2009wy}. To derive the extended first law, a new variation $\tilde{\delta}$ is introduced, which acts on the fields as
\begin{equation}
	\tilde{\delta}\Phi =\delta \Phi +\partial _{\Lambda}\Phi \tilde{\delta}\Lambda,
\end{equation}
incorporating variations of the cosmological constant. Accordingly, the extended version of Eq.~\eqref{eq_FirstLawGen} can be obtained as~\cite{Xiao:2023lap},
 \begin{equation}
 	\int_{\infty}\left[\tilde{\delta}\mathbf{Q}_{\xi_H} - \xi_H\cdot\mathbf{\Theta}[\Phi,\tilde{\delta}\Phi]\right] - \int_{\mathcal{H}}\tilde{\delta}\mathbf{Q}_{\xi_H} = \frac{\tilde{\delta}\Lambda}{8\pi}\int_{V_r}\xi_H\cdot\epsilon.
 	\label{eq_ExFirstLawGen}
 \end{equation}

\subsection{Three-Dimensional Case ($D=3$)}
We begin by investigating the rotating BTZ-like black hole solution. From the variation of the Lagrangian~\eqref{S1}, the presymplectic potential is given by
\begin{equation}
	\mathbf{\Theta}[\Phi,\delta\Phi]_{bc} = \left( 2{E_R}^{a\mu\nu\rho}\nabla_{\rho}\delta g_{\mu\nu} - 2\left(\nabla_{\rho}{E_R}^{a\mu\nu\rho}\right)\delta g_{\mu\nu} - \frac{1}{2}H^{a\mu\nu}\delta B_{\mu\nu} \right)\varepsilon_{abc},
\end{equation}
where
\begin{align}
	{E_R}^{abcd} &= \frac{1}{2\kappa}\left( X^{abcd} + \xi_2 B^{\mu\rho}{B^{\nu}}_{\rho}{Y_{\mu\nu}}^{abcd} + \xi_1 B^{\mu\nu}B_{\mu\nu}X^{abcd} \right), \\
	X^{abcd} &= g^{a[c}g^{d]b}, \\
	{Y_{\mu\nu}}^{abcd} &= \frac{1}{2}\left( {g_{(\mu}}^a{g_{\nu)}}^{[c}g^{d]b} - {g_{(\mu}}^b{g_{\nu)}}^{[c}g^{d]a} \right).
\end{align}
Accordingly, using Eqs. \eqref{eq:NoetherCurrent} and \eqref{eq_JdQ}, we obtain the Noether charge 
\begin{equation}
	(\mathbf{Q}_\xi)_c = \left( -{E_R}^{ab\mu\nu}\nabla_{\mu}\xi_{\nu} - 2\xi_{\mu}\nabla_{\nu}{E_R}^{ab\mu\nu} + \frac{1}{2}H^{ab\mu}B_{\mu\nu}\xi^{\nu} \right)\varepsilon_{abc}.
\end{equation}
For the three-dimensional rotating BTZ-like black hole, the Wald entropy is
\begin{equation}
	S_W = -2\pi \int_{S_h} {E_R}^{\mu\nu\rho\sigma}\epsilon_{\rho\sigma}\epsilon_{\mu\nu } = \frac{1}{2}(1+\ell)\pi r_h = \frac{1}{4}(1+\ell)A_h,
\end{equation}
where $\epsilon_{\mu\nu}$ denotes the binormal associated with the bifurcation surface. The corresponding Hawking temperature, derived from the surface gravity, is given by,
\begin{equation}
	T_H = \frac{\kappa}{2\pi} = \frac{r_h^4\Lambda_e - 16j^2}{2\pi r_h^3(1+\ell)},
\end{equation}
where $r_h$ denotes the horizon radius of the black hole. Using the Killing vector $\partial_t$ and $\partial_\phi$, we obtain from Eq.~\eqref{eq_deltaH}  the expressions for the energy and angular momentum of the black hole:
\begin{equation}
	E=\frac{1}{8}\left( 1+{\ell} \right) M, \quad J=\frac{j}{8}.
\end{equation}
The extended first law of black hole thermodynamics is given by Eq.~\eqref{eq_ExFirstLawGen}, from which we obtain
\begin{equation}
	\tilde{\delta}E + \frac{r^2}{8}\tilde{\delta}\Lambda - \Omega_h\tilde{\delta}J - T_H\tilde{\delta}S_W = \frac{r^2}{8}\tilde{\delta}\Lambda - \frac{\tilde{\delta}\Lambda}{8\pi}\pi r_h^2.
\end{equation}
Evidently, when $S_r$ is chosen as spatial infinity (i.e., $r\to\infty$), the coefficient of $\tilde{\delta}\Lambda$ exhibits a divergence. However, these divergent terms cancel rigorously, yielding the extended first law
\begin{equation}
	\tilde{\delta}E = T_H\tilde{\delta}S_W + \Omega_h\tilde{\delta}J - V\tilde{\delta} P,
\end{equation}
where the thermodynamic volume is  $V = \pi r_h^2$ and the pressure $P =-\Lambda/(8\pi)$. Using the scaling relations between the thermodynamic quantities, the Smarr relation can be obtained as
\begin{equation}
 T_H S_W + \Omega_h J - 2PV=0.
	\label{SM1}
\end{equation}

\subsection{Four-Dimensional Case ($D=4$)}
Similar to the three-dimensional case, the presymplectic potential and the Noether charge in four dimensions retain a similar structure,
\begin{align}
	\mathbf{\Theta}[\Phi,\delta\Phi]_{bcd} &= \left( 2{E_R}^{a\mu\nu\rho}\nabla_{\rho}\delta g_{\mu\nu} - 2\left(\nabla_{\rho}{E_R}^{a\mu\nu\rho}\right)\delta g_{\mu\nu} - \frac{1}{2}H^{a\mu\nu}\delta B_{\mu\nu} \right)\varepsilon_{abcd},\\
	(\mathbf{Q}_\xi)_{cd} &= \left( -{E_R}^{ab\mu\nu}\nabla_{\mu}\xi_{\nu} - 2\xi_{\mu}\nabla_{\nu}{E_R}^{ab\mu\nu} + \frac{1}{2}H^{ab\mu}B_{\mu\nu}\xi^{\nu} \right)\varepsilon_{abcd}.
\end{align}
The energy associated with the Killing vector $\partial_t$ is given by
\begin{equation}
	 E = (1+\ell_1) M.
\end{equation}
Moreover, the Wald entropy and the temperature are expressed as
\begin{equation}
	S_W = (1+\ell_1)\pi r_h^2 = \frac{1}{4}(1+\ell_1)A_h, \quad T_H = \frac{\kappa}{2\pi} = \frac{(1+\ell_1)^2 - (1+3\ell_1)r_h^2\Lambda}{4\pi(1+3\ell_1)(1+\ell_1)r_h}.
\end{equation}The formulation of extended thermodynamics yields
\begin{equation}
	\tilde{\delta}E = T_H\tilde{\delta}S_W + V\tilde{\delta} P, \quad E=2T_H S_W-2PV,
	\label{SM2}
\end{equation}
with the thermodynamic volume $V = 4\pi r_h^3/3$ and the pressure $P=-\Lambda/(8\pi)$.

Due to the introduction of the Kalb-Ramond field, both the area law of entropy and the gravitational energy receive small corrections. Comparing Eqs.~\eqref{SM1} and \eqref{SM2} with the corresponding results in general relativity ~\cite{Kastor:2009wy}, one can see that the Smarr relation remains unaltered. This is due to the fact that under the scaling transformation in thermodynamics, 
\begin{eqnarray}
E&\rightarrow&\lambda^{D-3}E  ,  \nonumber \\
	S&\rightarrow&\lambda^{D-2}S , \nonumber \\
	J&\rightarrow&\lambda^{D-2}J ,  \nonumber \\
	P&\rightarrow&\lambda^{-2}P ,  
\end{eqnarray}
the thermodynamic relations remain unchanged. In other words, the Lorentz symmetry breaking parameter enters the theory in a dimensionless manner.

We note that An~\cite{An:2024fzf} has studied black hole thermodynamics in bumblebee gravity using the Iyer-Wald formalism. Similar to gravity with a background Kalb-Ramond field, in bumblebee gravity, the spontaneous Lorentz violation arises from a potential $ V(B^\mu B_\mu) $ acting on a vector field $ B^\mu $.

The thermodynamic variables of the Schwarzschild-like 
bumblebee black hole is defined as~\cite{An:2024fzf}
\begin{equation}
	E = (1+\ell) M, S = (1+\ell)\pi r_h^2,	
\end{equation}
 and therefore the thermodynamic first law $dE=TdS$
holds.
But it was found that find the integral over the horizon,  $\int_{S_h} (\delta\textbf{Q}_{\xi_H} - \xi_H \cdot \mathbf{\Theta}[\Phi, \delta \Phi])$
has an additional contribution, which is absent in the standard Wald formalism. This additional contribution leads to a deviation in the entropy $S \neq S_W$, thus the Wald entropy does not satisfy the first law of thermodynamics in the bumblebee gravity model. 
The existence of the missing term relies on the divergent behavior of the bumblebee field at the horizon, namely, $B_r= |b| \frac{1+l}{1 - \frac{2M}{r}}$.

In the gravitational theory with the background Kalb-Ramond  field, the Kalb-Ramond field does not disperse at the event horizon, $B_{rt}=-B_{tr}= \frac{\sqrt{2}|b|}{2}$. The term $\int_{S_h} (\delta\textbf{Q}_{\xi_H} - \xi_H \cdot \mathbf{\Theta}[\Phi, \delta \Phi])$
will have no additional contribution, thus the Wald entropy satisfies the first law of thermodynamics.

\section{ CONCLUSIONS }
In this work,  we studied black hole solutions in the
presence of a nonzero VEV of Kalb-Ramond field in the three-dimensional stationary axisymmetric spacetime and the four-dimensional static spherically 
symmetric spacetime. We presented exact solutions for 
four-dimensional static spherically symmetric black holes 
and three-dimensional rotating black holes.

First, we noted that in some previous works, the variation of the nonminimal coupling term between the Kalb-Ramond field and the scalar curvature was neglected. Instead, this term was absorbed into a redefinition of variables. We have properly taken this term into account and obtained new four-dimensional spherically symmetric black hole solutions. These solutions modify the previous results  given in Ref.~\cite{Yang:2023wtu}, where the cosmological constant was $ \Lambda = 0 $ for the potential  $ V(X) = \frac{\lambda}{2}X^2 $, and $ \Lambda \neq 0 $ for  $ V(X) = \frac{\lambda}{2}X $. Additionally, by incorporating this term, we have also obtained new black hole solutions in the case of a nonzero cosmological constant for the potential  $ V(X) = \frac{\lambda}{2}X^2 $. This allows for a direct consideration of black hole solutions  with and without a cosmological constant under the potential $ V(X) = \frac{\lambda}{2}X^2 $.

Moreover, we obtained a rotating BTZ-like black hole solution. When $ \ell = 0 $, this solution is consistent with the rotating BTZ black hole solution. When $ j = 0 $, the solution reduces to a static BTZ-like black hole.

Finally we studied the thermodynamics of black holes in gravity with a background Kalb-Ramond field using Iyer-Wald formalism. We  derived the first law of black hole thermodynamics. Due to the introduction of the Kalb-Ramond field, both the area law of entropy and the gravitational energy receive small corrections, while the Smarr relation remains unaltered.  Our analysis further revealed that the Kalb-Ramond field remains non-dispersive at the event horizon. Consequently, the Wald entropy is shown to satisfy the first law of thermodynamics, a behavior distinct from that observed in bumblebee gravity black holes within the Iyer-Wald covariant phase space formalism.

In summary, we have obtained new classes of black hole solutions in gravity with a background Kalb–Ramond field and analyzed their thermodynamic properties using the Iyer–Wald formalism. These solutions, characterized by Lorentz-violating parameters, introduce small but potentially observable deviations from general relativity. In the absence of a cosmological constant, the solution exhibits slight departures from asymptotic flat spacetime. These solutions provide a theoretical framework to probe Lorentz-symmetry violations in strong-gravity regimes. In particular, possible observational signatures—such as quasinormal modes, black hole shadows, and extreme mass-ratio inspirals—may offer promising avenues to test the physical relevance of these effects.

\section*{  Acknowledgements }
	This work was supported by 
	the National Key Research and Development Program of China (Grant No. 2020YFC2201503), 
	the National Natural Science Foundation of China (Grants No. 12475056, No. 12247101, and No. 12475055 ), 
	the 111 Project (Grant No. B20063), 
	Gansu Province's Top Leading Talent Support Plan.

\bibliography{REF}

\end{document}